\documentclass[aps,prd,onecolumn,nofootinbib,superscriptaddress]{revtex4}
\pdfoutput=1
\usepackage{float}
\usepackage{graphicx}
\usepackage{amsmath}
\usepackage{amsfonts}
\usepackage{amssymb,ulem}
\usepackage{mathrsfs}
\usepackage{color}%
\usepackage{dcolumn}
\usepackage{subfigure}
\usepackage{pdfpages}
\usepackage{multirow}

\newcommand{\dd}{\mathrm{d}}

\newcommand{\fsp}{f_{\mathrm{sp}}}
\newcommand{\Mh}{\mathcal M_{\mathrm D}}
\newcommand{\MK}{M_{\mathrm K}}
\newcommand{\Deltaom}{\Delta\omega}
\usepackage{MnSymbol,wasysym}
\usepackage{braket,diagbox}
\usepackage{eurosym}
\usepackage{euscript}
\usepackage[usenames,dvipsnames,svgnames]{xcolor}

\newcommand{\RNum}[1]{\uppercase\expandafter{\romannumeral #1\relax}}
\usepackage[colorlinks=true,linkcolor=blue,urlcolor=blue,filecolor=black,citecolor=red,
pdfstartview=FitV,pdftitle={},pdfsubject={},pdfkeywords={},pdfpagemode=None,bookmarksopen=true]{hyperref}

\begin{document}

\baselineskip=0.5 cm
\title{Constraining  Dehnen-inspired dark matter halo with S2 star's orbit}

\author{Qi Qi}
\email{qiqiphy@163.com}
\affiliation{Center for Gravitation and Cosmology, College of Physical Science and Technology, Yangzhou University, Yangzhou, 225009, China}

\author{Qiang Wu}
\email{wuq@zjut.edu.cn}
\affiliation{Institute for Theoretical Physics and Cosmology,
Zhejiang University of Technology, Hangzhou 310023, China}

\author{Xiao-Mei Kuang}
\email{xmeikuang@yzu.edu.cn (corresponding author)}
\affiliation{Center for Gravitation and Cosmology, College of Physical Science and Technology, Yangzhou University, Yangzhou, 225009, China}


\begin{abstract}
\baselineskip=0.5 cm
We use the orbit of S2 star around Sagittarius A* to constrain three black hole backgrounds motivated by Dehnen profiles with inner slopes $\gamma=0$, $1$, and $5/2$, which are characterized by the scale radius $r_s$ and density $\rho_s$ of the dark matter halo.  An exact turning-point integral gives the pericenter advance at finite eccentricity, and an exterior expansion identifies the leading halo contributions. A fifteen-parameter Bayesian analysis combines astrometric and spectroscopic observations with the measured precession parameter, from which we obtain the marginalized $1\sigma$ upper limits $r_s/M<0.236$, $0.168$, and $0.030$, and $\rho_sM^2<0.536$, $0.578$, and $0.004$, respectively. The leading halo monopole in the first two metrics is degenerate with the central mass. At fixed dynamical mass, their leading radial profile correction produces retrograde precession. The $\gamma=5/2$ metric introduces an additional angular deficit contribution, which enhances the precession response and accounts for its smaller parameter limits. 
\end{abstract}

\maketitle
\tableofcontents
\newpage

\section{Introduction}

Supermassive black holes (SMBHs) reside at the centers of many galaxies \cite{Gultekin:2009qn,Genzel:2010zy}. The Event Horizon Telescope (EHT) has obtained horizon-scale images of M87* and Sagittarius A* (Sgr~A*), the compact radio source at the Galactic Center \cite{EventHorizonTelescope:2019dse,EventHorizonTelescope:2022wkp}. These images provide electromagnetic evidence for the central black holes and resolve the surrounding emission structures. Comparisons with relativistic accretion models constrain the near-horizon spacetime and accretion flow, enabling tests of general relativity (GR) \cite{Bambi:2019tjh,EventHorizonTelescope:2019pgp,EventHorizonTelescope:2019ggy,EventHorizonTelescope:2020qrl,EventHorizonTelescope:2021srq,EventHorizonTelescope:2021dqv,EventHorizonTelescope:2022xqj}. Gravitational waves offer a complementary probe of massive black-hole systems. These systems are among the targets of planned space-based interferometers, including the Laser Interferometer Space Antenna (LISA) \cite{LISA:2017pwj}, Taiji \cite{Hu:2017mde}, and TianQin \cite{TianQin:2015yph}.

Stellar orbits around Sgr~A* probe its gravitational field at larger radii. Near-infrared monitoring at the Keck Observatory \cite{Ghez:2003qj} and the Very Large Telescope (VLT) \cite{Gillessen:2017jxc} has traced the motions of the surrounding S-stars for more than two decades. Interferometric astrometry with GRAVITY on the Very Large Telescope Interferometer (VLTI) has further improved their orbital measurements \cite{GRAVITY:2021xju,Dayem:2026ktt}. S2 follows a highly eccentric orbit ($e\simeq0.88$), with a period of about 16 years and a pericenter distance of approximately $120~\mathrm{AU}$. Spectroscopic observations near its 2018 pericenter passage detected the combined gravitational and transverse Doppler redshift \cite{GRAVITY:2018ofz,Do:2019txf}. Subsequent astrometric measurements resolved the Schwarzschild precession of its orbit \cite{GRAVITY:2020gka,GRAVITY:2021xju}. Both effects are consistent with GR within the measurement uncertainties. The precision of S2 astrometry and spectroscopy also allows tests of departures from GR \cite{Shaymatov:2023jfa,DellaMonica:2021xcf,DellaMonica:2022eeg,Li:2022rjv,Yan:2022fkr,Zhang:2024fpm,deMartino:2021daj,Li:2024tld,Fernandez:2023kro,Li:2025qcv,GRAVITY:2023azi,GRAVITY:2023cjt,DellaMonica:2023dcw,QiQi:2024dwc,Wu:2025xtn,QiQi:2026pnb,Qi:2026npq,Zhao:2026hfl,Fang:2026jtw}.

Dark matter can contribute to the extended mass around Sgr~A* through a smooth halo, a density spike, or a coherent field configuration \cite{Buckley:2017ijx}. The Navarro--Frenk--White and Burkert profiles describe cusped and cored halos, respectively \cite{Navarro:1996gj,Burkert:1995yz}. Black-hole growth, stellar heating, and particle capture can reshape these distributions near the central black hole \cite{Gondolo:1999ef,Merritt:2003qk,Sadeghian:2013laa}. 
The Dehnen family consists of spherical, finite-mass density profiles with an adjustable inner slope and an outer $r^{-4}$ falloff \cite{Dehnen:1993uh}. It includes the Hernquist and Jaffe profiles \cite{Hernquist:1990be,Jaffe:1983iv} and belongs to the broader class of double-power-law models \cite{Lilley:2018mly}. Black-hole metrics constructed from Dehnen potentials or circular-velocity relations have been used to study shadows, geodesics, and perturbations for several inner slopes \cite{Gohain:2024eer,Jha:2024ltc,Jha:2025xjf,Al-Badawi:2024asn,Liang:2025vux}. A relativistic halo model relates the density and matter stresses to the two independent metric functions of a static, spherically symmetric spacetime \cite{Cardoso:2021wlq}. The mapping from halo parameters to orbital observables therefore depends on the metric construction as well as the density profile.

Analyzes of S-star orbits have constrained dark-matter spikes \cite{Lacroix:2018zmg,Nampalliwar:2021tyz,Shen:2023kkm}, ultralight bosonic dark matter \cite{DellaMonica:2023dcw,Yuan:2022nmu,DellaMonica:2022kow}, and scalar and vector clouds \cite{GRAVITY:2023cjt,GRAVITY:2023azi}. The inferred bounds depend on the assumed radial distribution and its characteristic scale. Recent work has used shadow observables and S2 orbits to constrain generalized halo profiles \cite{Li:2025qcv}. Related studies have fitted a generalized Dehnen mass-function metric to S2 data \cite{Xamidov:2026dps} and compared the mass precession of different S-stars \cite{Paul:2026ikn}. For the potential-based Dehnen metrics considered here, orbital deviations can arise from changes in the effective central mass, the radial profile, and the asymptotic geometry. Separating these contributions identifies the parameter combinations to which the S2 data are sensitive.

In this work, we study three Dehnen-inspired black-hole metrics labeled by the inner slopes $\gamma=0$, $1$, and $5/2$. We derive their pericenter advance from the orbital turning points without expanding in eccentricity. A common exterior expansion separates the monopole mass shift, the leading radial-profile correction, and the angular-deficit contribution. We then jointly fit the orbital, reference-frame, and metric parameters to the S2 data. The remaining of this paper is organized as follows. In section~\ref{sec:dynamics}, starting from the timelike geodesic, we derive its pericenter precession in three black hole metrics, and then give the preliminary parameter regime. In section~\ref{sec:data}, we describe the available astrometric and spectroscopic data used in this paper, and then we construct the orbital model, which will be 
used in the MCMC simulation. We then perform the MCMC study and analyze the results in section~ \ref{sec:results}. The last section contributes to the conclusions and discussions.
We use geometric units $G=c=1$ and metric signature $(-+++)$ in the dynamical calculation, restoring physical units for the observables.

\section{Geodesic motion and pericenter precession}\label{sec:dynamics}

In this section, we first briefly introduce the black hole motivated by Dehnen profile with $\gamma=0,~1$, and $5/2$. Then we derive the geodesic equations for timelike test particles and calculate the correction to the pericenter precession ratio introduced by the dark matter halo.

\subsection{Dehnen profiles and metric families}

The Dehnen density and enclosed mass are \cite{mo2010galaxy}
\begin{eqnarray}
 \rho_{\mathrm D}(r)&=&\rho_s\left(\frac{r}{r_s}\right)^{-\gamma}
 \left(1+\frac{r}{r_s}\right)^{\gamma-4},\label{eq:dehnen}\\
 M_{\mathrm D}(<r)&=&\Mh\left(\frac{r}{r+r_s}\right)^{3-\gamma},
 \qquad \Mh=\frac{4\pi\rho_s r_s^3}{3-\gamma},\label{eq:enclosed}
\end{eqnarray}
with $0\leq\gamma<3$. The parameters $r_s$ and $\rho_s$ are the scale
radius and scale density. The density is finite at the centre only for
$\gamma=0$. The nominal halo mass $\Mh$ is distinct from the central
metric parameter $M$.

The three black hole metrics we consider here take the form
\begin{eqnarray}
 \dd s^2=-f_\gamma(r)\dd t^2+\frac{\dd r^2}{f_\gamma(r)}
             +r^2\dd\Omega^2,~~\text{with}~~  f_\gamma(r)=1-\frac{2M}{r}-h_\gamma(r) \label{eq:metric},
\end{eqnarray}
where \cite{Gohain:2024eer,Jha:2025xjf,Al-Badawi:2024asn}
\begin{eqnarray}
 h_0(r)&=&\frac{4\pi\rho_s r_s^3(r_s+2r)}{3(r+r_s)^2},\label{eq:h0}\\
 h_1(r)&=&\frac{4\pi\rho_s r_s^3}{r+r_s},\label{eq:h1}\\
 h_{5/2}(r)&=&32\pi\rho_s r_s^2\sqrt{1+\frac{r_s}{r}}.\label{eq:h25}
\end{eqnarray}
Each metric reduces to Schwarzschild metric when $\rho_s=0$ or $r_s=0$.
The label $\gamma$ identifies the generating Dehnen profile.

The Newtonian potential normalized at infinity is
\begin{equation}
 \Phi_{\mathrm D}(r)=-\frac{\Mh}{r_s(2-\gamma)}
 \left[1-\left(\frac{r}{r+r_s}\right)^{2-\gamma}\right],\label{eq:potential}
\end{equation}
for $\gamma\ne2$. The first two functions satisfy
$h_\gamma=-2\Phi_{\mathrm D}$ and the third satisfies
$h_{5/2}=-2\Phi_{\mathrm D}+\mathcal A$, where
\begin{equation}
 \mathcal A=32\pi\rho_s r_s^2.\label{eq:deficit}
\end{equation}
All three reproduce the generating Newtonian radial force,
\begin{equation}
 -\frac{r^2 h_\gamma'(r)}{2}=M_{\mathrm D}(<r).\label{eq:force}
\end{equation}
The choice $g_{rr}=f_\gamma^{-1}$ also fixes the relativistic stress
tensor. With $m_{\mathrm E}=M+rh_\gamma/2$ and considering that energy-momentum tensor of source $T^{\mu}_{\nu}=\text{diag}(-\rho_{\mathrm E},p_r,p_t,p_t)$, the Einstein's equations give
\begin{equation}
 \rho_{\mathrm E}=\frac{m_{\mathrm E}'}{4\pi r^2},\qquad
 p_r=-\rho_{\mathrm E},\qquad
 p_t=-\frac{m_{\mathrm E}''}{8\pi r}.\label{eq:stress}
\end{equation}
It is noted that these source densities differ from Eq.~\eqref{eq:dehnen}, as shown in
Appendix~\ref{app:source}. 

It is straightforward to see that the first two metrics are asymptotically flat. The third has
\begin{equation}
 f_\infty=\kappa\equiv1-\mathcal A.\label{eq:kappa}
\end{equation}
A static asymptotic region has $\kappa>0$. The time coordinate
$T=\sqrt\kappa\,t$ normalizes the asymptotic lapse. The spatial metric
remains $\dd r^2/\kappa+r^2\dd\Omega^2$, with a nonzero angular deficit.

\subsection{Exact orbital integral and precession ratio}
Then, we follow the procedure of \cite{Drasco:2003ky} to derive the correction to pericenter precession comparing to the Schwarzschild case.
Since the black holes we consider have spherical symmetry, we can place the orbit in the equatorial plane with $\theta=\pi/2$ for simplicity.
The specific Killing energy and angular momentum are 
\begin{equation}
 E=f\dot t,\qquad L=r^2\dot\varphi,\label{eq:EL}
\end{equation}
where an overdot denotes the differentiation on  proper-time. Then, the timelike
normalization gives us
\begin{equation}
 \dot r^2=\mathcal R(r)
 =E^2-f(r)\left(1+\frac{L^2}{r^2}\right).\label{eq:radial}
\end{equation}
The pericenter and apocentre define the orbital labels
\begin{equation}
 r_p=a(1-e),\quad r_a=a(1+e),\quad
 p=a(1-e^2).\label{eq:turning}
\end{equation}
The conditions $\mathcal R(r_p)=\mathcal R(r_a)=0$ determine
\begin{eqnarray}
 L^2&=\frac{r_p^2r_a^2[f(r_a)-f(r_p)]}
              {r_a^2f(r_p)-r_p^2f(r_a)},\label{eq:Lturn}~~~
 E^2&=\frac{f(r_p)f(r_a)(r_a^2-r_p^2)}
              {r_a^2f(r_p)-r_p^2f(r_a)}.\label{eq:Eturn}
\end{eqnarray}

Subsequently, the pericenter advance per radial period is \cite{schmidt2002celestial}
\begin{equation}
 \Deltaom=2\int_{r_p}^{r_a}
       \frac{L\,\dd r}{r^2\sqrt{\mathcal R(r)}}-2\pi.\label{eq:apsidal}
\end{equation}
To evaluate the pericenter, it is convenient to introduce the radial anomaly $\chi$ defined by
\begin{equation}
 r(\chi)=\frac{p}{1+e\cos\chi},\label{eq:darwin}
\end{equation}
while will give the regular half-orbit integral
\begin{equation}
 \frac{\dd\varphi}{\dd\chi}
 =\frac{Le\sin\chi}{p\sqrt{\mathcal R[r(\chi)]}},
 \qquad 0<\chi<\pi.\label{eq:darwinintegrand}
\end{equation}
Its endpoint limits are finite for nondegenerate turning points.
This radial parameter agrees with the true anomaly in the Kepler limit.
In Schwarzschild spacetime,
\begin{equation}
 \frac{\dd\varphi}{\dd\chi}
 =\left[1-\frac{2M}{p}(3+e\cos\chi)\right]^{-1/2},\label{eq:schwexact}
\end{equation}
which yields
\begin{equation} \label{eq:dwGR}
\Deltaom_{\rm GR}=\frac{6\pi M}{p}.
\end{equation}

At $r_s\ll r_p$, the three lapses in Eq. \eqref{eq:metric} share the expansion
\begin{equation}
 \begin{gathered}
 f_{\gamma}(r)=\kappa-\frac{2\mu}{r}+\frac{Q}{r^2}+O(r^{-3}),\\
 \mu=M+\Mh,\qquad Q=4\pi\rho_s r_s^4.
 \end{gathered}\label{eq:outer}
\end{equation}
Here $\kappa=1$ for $\gamma=0,1$ and is given by
Eq.~\eqref{eq:kappa} for $\gamma=5/2$. We then have 
\begin{equation}
 L^2=\frac{\mu p-Q}
 {\kappa-\mu(3+e^2)/p+2Q(1+e^2)/p^2}.\label{eq:Louter}
\end{equation}
Further factoring the radial polynomial gives Eq.\eqref{eq:darwinintegrand} as 
\begin{eqnarray}
 \left(\frac{\dd\varphi}{\dd\chi}\right)^{-2}=\kappa+\frac{Q}{L^2}-\frac{2\mu}{p}(3+e\cos\chi)+\frac{Q}{p^2}\left[6+4e\cos\chi+e^2(1+\cos^2\chi)\right].
 \label{eq:factorization}
\end{eqnarray}
Subsequently, at fixed $\kappa>0$, in the approximate case   with $\mu/(\kappa p)\ll1$ and
$Q/(\mu p)\ll1$, we can evaluate pericenter advance per radial period  in Eq.\eqref{eq:apsidal} in general formula as 
\begin{equation}
 \Deltaom=2\pi(\kappa^{-1/2}-1)
 +\frac{6\pi\mu}{\kappa^{3/2}p}
 -\frac{\pi Q}{\mu\sqrt\kappa\,p}+O(p^{-2}).\label{eq:unified}
\end{equation}
It is worthwhile to point out that the three terms describe the angular deficit, the monopole contribution,
and the leading radial profile correction, respectively.

\begin{figure} [h]
{\centering
\includegraphics[width=2.30in]{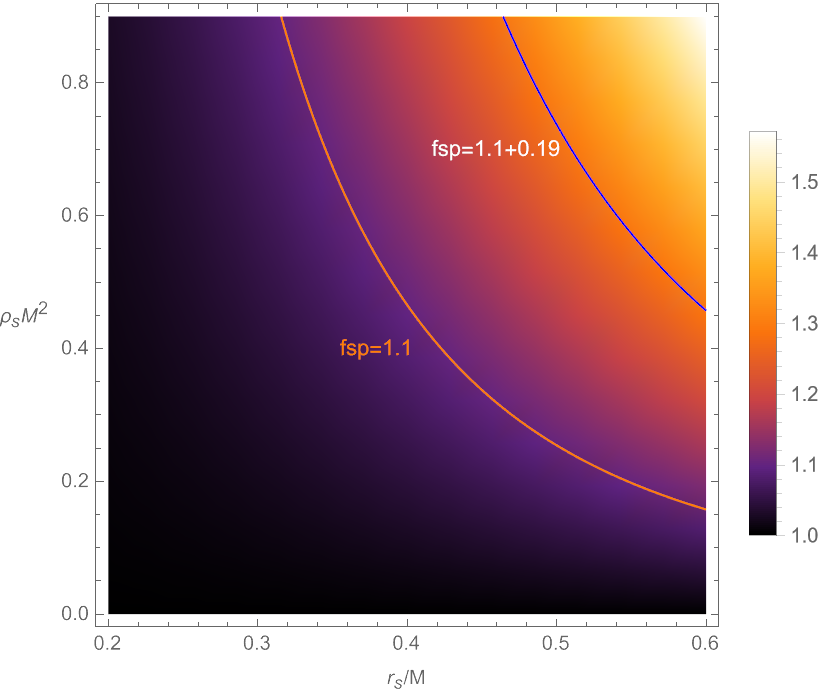}\hspace{0.1cm}
\includegraphics[width=2.30in]{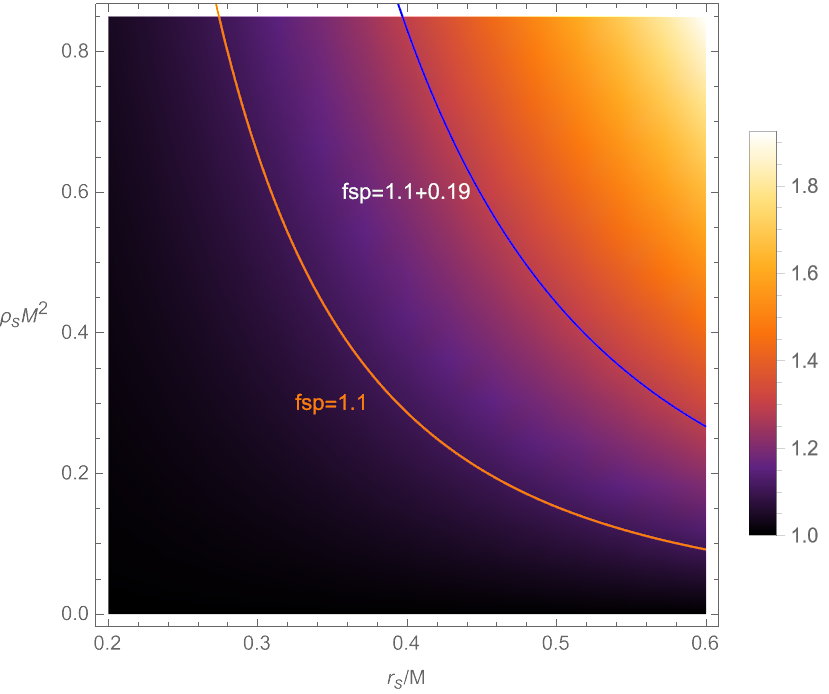}\hspace{0.1cm}
\includegraphics[width=2.30in]{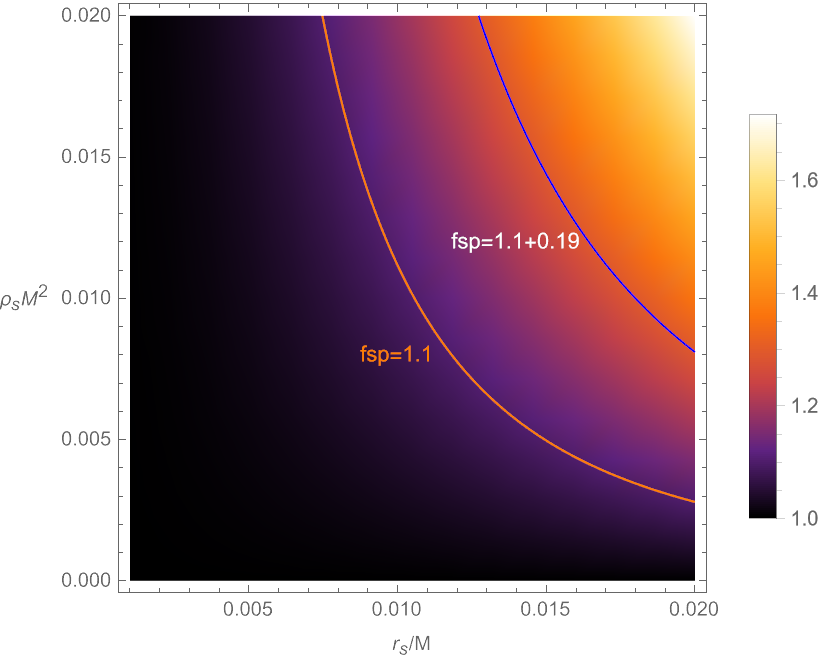}
   \caption{Density plot of $\fsp$ for the S2 star in the $(r_s/M,\rho_sM^{2})$ plane, assuming the central SMBH Sgr A* is surrounded by a Dehnen-type dark matter halo. From left panel to the right, the plots indicate the results for the black hole embedded in the Dehnen-$(1,4,0)$,  $(1,4,1)$ and $(1,4,5/2)$ halos, respectively. In the calculation, we set $a=125.058\,\mathrm{mas}$, $e=0.88465$ and $M=4.26\times10^{6}\,M_{\odot}$. } \label{fig:constrainf_sp}}
\end{figure}

Considering the dimensionless parameters are
$r_0=\frac{r_s}{M}$ and $\rho_0=\rho_sM^2$, 
we then linearize Eq.~\eqref{eq:unified} in the halo amplitudes at fixed $M$, of which the leading orders are 
\begin{eqnarray}
 \Deltaom_0&=&\frac{6\pi M}{p}
 +\frac{4\pi^2\rho_0r_0^3(2-r_0)M}{p}+\cdots,\label{eq:dw0}\\
 \Deltaom_1&=&\frac{6\pi M}{p}
 +\frac{4\pi^2\rho_0r_0^3(3-r_0)M}{p}+\cdots,\label{eq:dw1}\\
 \Deltaom_{5/2}&=&\frac{6\pi M}{p}+32\pi^2\rho_0r_0^2+\frac{4\pi^2\rho_0r_0^2(72+12r_0-r_0^2)M}{p}+\cdots.
 \label{eq:dw25}
\end{eqnarray}
for $\gamma=0,~1$, and $5/2$, respectively. The dots denote subleading terms. It is noted that this additional expansion requires $\Mh/M\ll1$ and, for the third
branch, $\mathcal A\ll1$. The compact scale condition remains
$r_s/r_p\ll1$.
Further combining Eqs.~\eqref{eq:dw0}--\eqref{eq:dw25} with Eq.~\eqref{eq:dwGR} gives the model predictions of the precession ratio,
\begin{eqnarray}
 \fsp^{(0)}&=&1+\frac{2\pi}{3}\rho_0r_0^3(2-r_0),\label{eq:fsp0}\\
 \fsp^{(1)}&=&1+\frac{2\pi}{3}\rho_0r_0^3(3-r_0),\label{eq:fsp1}\\
 \fsp^{(5/2)}&=&1+\frac{16\pi p}{3M}\rho_0r_0^2+\frac{2\pi}{3}\rho_0r_0^2(72+12r_0-r_0^2).
 \label{eq:fsp25}
\end{eqnarray}

The best-fit ratio of the measured precession to the GR prediction is given by \cite{GRAVITY:2020gka}
\begin{eqnarray}\label{eq:fsp}
\fsp\equiv \frac{\Delta \omega_{\rm s2}}{\Delta \omega_{\rm GR}}=1.1\pm 0.19,
\end{eqnarray}
within $1\sigma$ uncertainty which leaves some room for a surrounding dark matter halo or alternative gravity beyond GR.
The measured $\fsp$ in the $(r_s/M,\rho_sM^{2})$ plane corresponding to the three models are illustrated in Fig.~\ref{fig:constrainf_sp}, in which the orange curve denotes the best value  while the blue curve denotes the upper $1\sigma$ uncertainty. Therefore  the parameter region  below the blue curve could be available within $1\sigma$ uncertainty.

\section{S2 data and observables}
\label{sec:data}

The S2 star moves on the shortest orbit known around the radio source Sgr A*. Its period is about 16 years, its semi-major axis is roughly $970\,$AU, and its eccentricity reaches $e\simeq0.88$, so the star passes within some $120\,$AU of the central mass at every pericenter passage. Three decades of monitoring by the GRAVITY Collaboration have turned this single star into a precisely tracked object in the Galactic Centre: the astrometric positions and the radial velocities are now good enough that the relativistic corrections themselves are measured, and both the gravitational redshift and the periastron advance $\delta\varphi\simeq12'$ per revolution have been detected. All of these measurements are consistent with GR. The trajectory of the star is fixed by the geometry of the spacetime it travels through, so the distribution of matter around the hole bends the orbit in a way that is distinguishable from the vacuum Schwarzschild geometry. In this section
we take the three Dehnen-type halo backgrounds of section~\ref{sec:dynamics} and use the public S2 data to constrain the dark matter halo parameters $(r_s,\rho_s)$.

\subsection{S2 star data}

Our MCMC fit employs three types of observables. First, the astrometric positions determine the size, shape, and orientation of the orbit projected onto the sky. Second, the radial velocities describe the motion along the line of sight, where the relativistic frequency shift plays an important role. Third, the measured periastron advance directly constrains the orbital precession. These three observables are described below.

\begin{figure} [h]
{\centering
\includegraphics[width=3.2in]{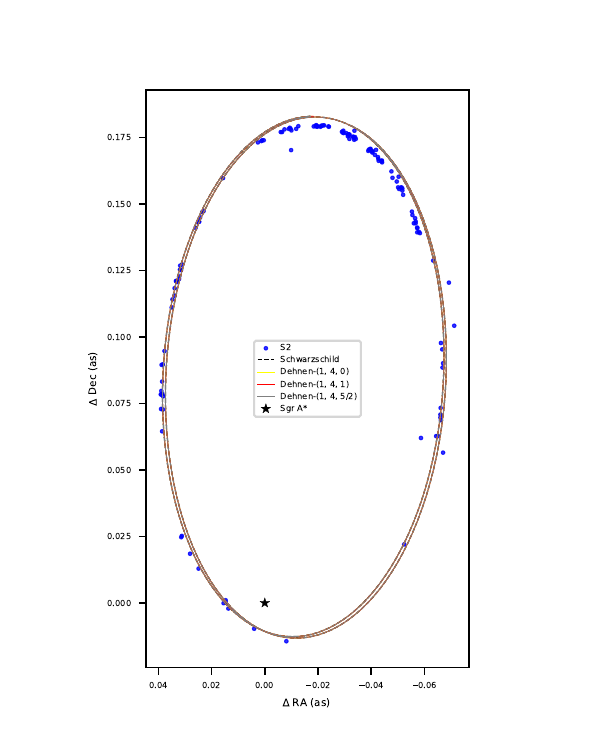}
\includegraphics[width=3.8in]{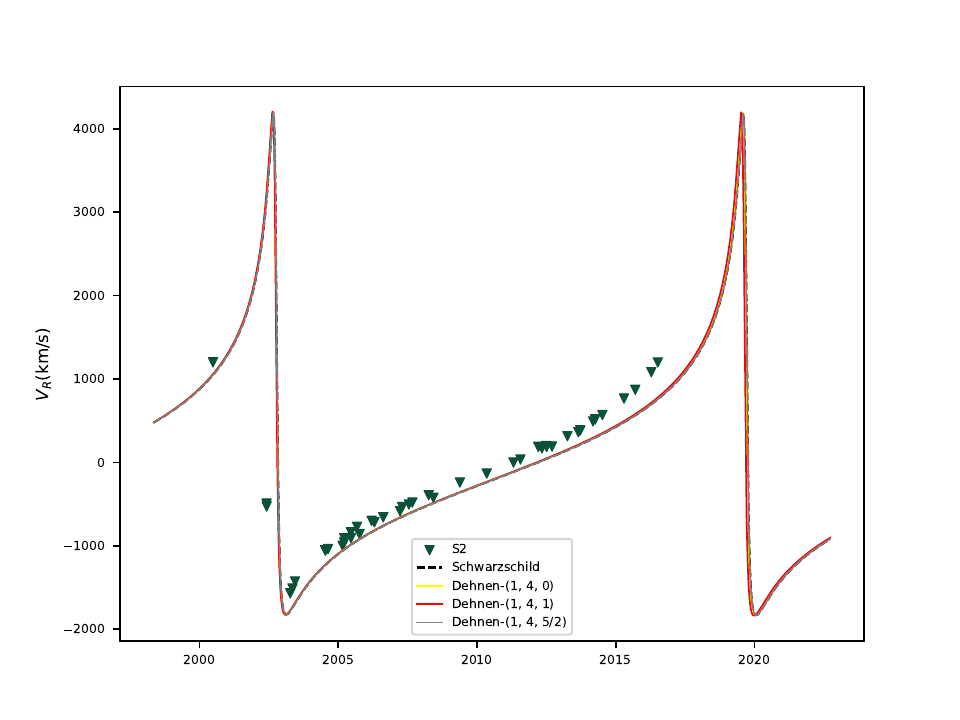}
\caption{{\bf Left:} Comparison of the calculated orbit and the measured
astrometric positions of the S2 star. The black dashed line represents the
Schwarzschild black hole prediction, while the yellow, red, and gray solid
lines show the results for $\gamma=0,1,$ and $5/2$ respectively. Blue dots indicate the observed
positions, and Sgr~A* is positioned at the origin, marked by a black star.
{\bf Right:} Radial velocity of the S2 star as a function of time. Similarly, the black dashed curve corresponds to the Schwarzschild prediction, while the yellow, red, and gray solid curves represent the three Dehnen models. Green inverted triangles show the observed radial velocity data.}
\label{fig:OBC} }
\end{figure}

{\bf Astrometric positions (AP):} This sample contains 145 sky positions of S2 covering 1992.224--2016.53 \cite{Gillessen:2017jxc}. Epochs before 2002 were obtained with the speckle-imaging camera SHARP at the ESO New Technology Telescope (NTT), at a typical accuracy of about $3.8\,$mas. From 2002 onwards the star was followed with the adaptive-optics camera NAOS+CONICA (NACO) at the Very Large Telescope (VLT), which improved the precision to roughly $400\,\mu$as. These measurements are the blue dots in the left panel of Fig.~\ref{fig:OBC}.

{\bf Radial velocities ($V_R$):} This sample consists of 44 measurements between 2000.487 and 2016.519 \cite{Gillessen:2017jxc}, again divided between two instruments: the NIRC2 spectrograph at the Keck Observatory for the early epochs, and the adaptive-optics integral-field spectrograph SINFONI at the VLT from 2003 onwards. They are plotted as the green inverted triangles in the right panel of Fig.~\ref{fig:OBC}.

{\bf Orbital precession ($\Delta \phi$):} In addition to the astrometric and radial velocity data, we include the pericenter-advance measurement of S2 reported by the GRAVITY Collaboration \cite{GRAVITY:2020gka}. The measured precession ratio is given by Eq.~\eqref{eq:fsp} and it agrees with the GR prediction within $1\sigma$ and rules out a purely Newtonian orbit at more than $5\sigma$. Since the full GRAVITY astrometric solution has not been released, we adopt this ratio as the third data component. On the theoretical side, the precession is given by Eqs.~\eqref{eq:fsp0}--\eqref{eq:fsp25}.

\subsection{From orbit to observables }

\begin{figure} [h]
{\centering
\includegraphics[width=2.5in]{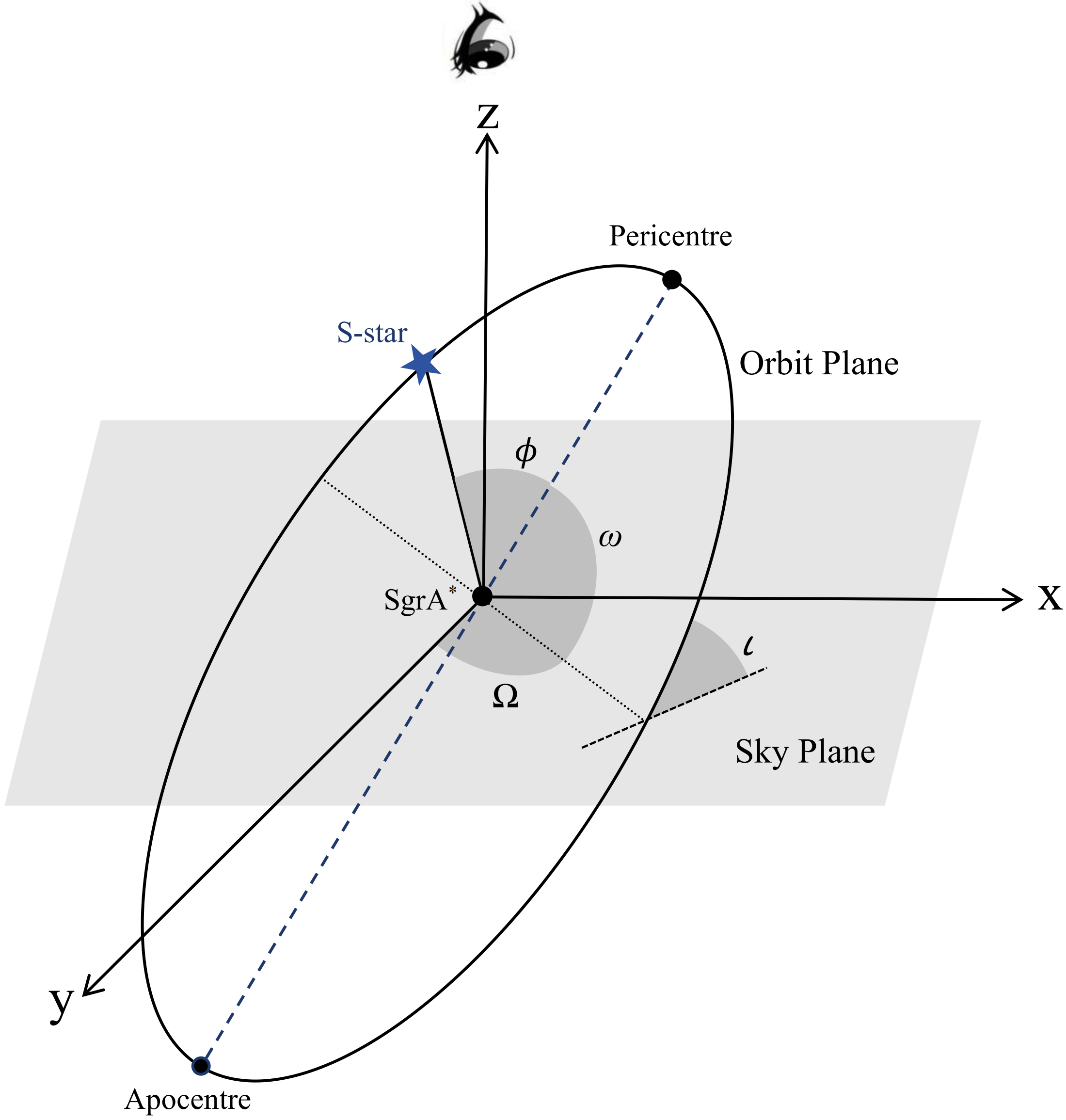} \hspace{0.5cm}
   \caption{The relationship between the orbital plane of the S-star and the observer's plane, along with the relevant Keplerian orbital elements. Here, the $Z$-axis coincides with the observer's line of sight, $\Omega$ is the longitude of ascending node, $\omega$ is the argument of pericenter (i.e., the angle from the ascending node to the pericenter measured in the orbital plane), the pentagram is the position of the S-star, and $\iota$ (the inclination) is the angle between the observer's plane and the orbital plane. }   \label{fig:CT}}
\end{figure}

Parameter estimation needs the full trajectory, so we integrate the
geodesic equations numerically, with the halo entering through the lapse
function Eq.~\eqref{eq:metric}. The observations, however, refer to the sky plane of the observer and to the line of sight, so the two frames have to be related before any comparison with data is possible, the relationship between them is shown in Fig.~\ref{fig:CT}. This transformation is conventionally written with the Thiele-Innes constants,
\begin{eqnarray}
X &= B\,x_{\rm orb}+G\,y_{\rm orb}, \label{eq:X}\\
Y &= A\,x_{\rm orb}+F\,y_{\rm orb}, \label{eq:Y}\\
Z &= C\,x_{\rm orb}+H\,y_{\rm orb}, \label{eq:Z}
\end{eqnarray}
where $(x_{\rm orb},y_{\rm orb},z_{\rm orb})$ are the Cartesian coordinates in the orbital plane and $(X,Y,Z)$ those in the sky frame, $Z$ being counted along the line of sight towards the observer. The Thiele–Innes
coefficients are
\begin{eqnarray}
A &=&\cos\Omega\cos\omega-\sin\Omega\sin\omega\cos\iota, \label{eq:A}\nonumber\\
B &=&\sin\Omega\cos\omega+\cos\Omega\sin\omega\cos\iota, \label{eq:B}\nonumber\\
F &=&-\cos\Omega\sin\omega-\sin\Omega\cos\omega\cos\iota, \label{eq:F}\nonumber\\
G &=&-\sin\Omega\sin\omega+\cos\Omega\cos\omega\cos\iota, \label{eq:G}\nonumber\\
C &=&-\sin\omega\sin\iota, ~~~~H=-\cos\omega\sin\iota, \label{eq:H}
\end{eqnarray}
and are fixed by the three orbital elements above. The same rotation acts on velocities. Recession has the positive sign
$v_{\rm los}=-\dd Z/\dd T$, where $\dd T=\sqrt{f(r_{\rm em})}\,\dd t$
is the observer-normalized time and the derivative is evaluated at emission.

 Also, the measured astrometric positions refer to the photon arrival time $t_{\rm obs}$, whereas the orbital solution is parametrized by the emission time $t_{\rm em}$. The finite light-travel time introduces the R\o mer delay. After absorbing the constant propagation time to the Galactic Center into the time origin, the leading correction reads \cite{Do:2019txf,GRAVITY:2018ofz}
\begin{equation}
t_{\rm em} = t_{\rm obs} - \frac{Z(t_{\rm em})}{c},
\end{equation}
where $Z(t_{\rm em})$ is the line-of-sight coordinate given by Eq.~\eqref{eq:Z}, and $c$ is the speed of light. Solving this implicit relation at each observational epoch determines the emission time at which the sky-projected orbital position should be evaluated.

A further correction concerns the reference frame itself, which is not
perfectly centred on Sgr A*: constant offsets and slow drifts of the
astrometric solution are absorbed by four other parameters,
\begin{eqnarray}
X &=& X(t_{\rm em})+x_0
     +v_{x0}\,(t_{\rm em}-t_{\rm refer}), \label{eq:Xcorr}\\
Y &=& Y(t_{\rm em})+y_0
     +v_{y0}\,(t_{\rm em}-t_{\rm refer}). \label{eq:Ycorr}
\end{eqnarray}
Here $t_{\rm refer}$ denotes the reference epoch associated with the astrometric solution and we will adopt  $t_{\rm refer}=2009.02$ \cite{plewa2015pinpointing}. $x_0$ and $y_0$ are
constant shifts, while $v_{x0}$ and $v_{y0}$ describe a
linear drift of the frame relative to the dynamical centre. It receives a special-relativistic part from the motion of the star and a general-relativistic part from the potential at the point of emission, called Doppler shift and gravitational redshift,
\begin{eqnarray}
\zeta_D &=&\frac{\sqrt{1-v_{\rm em}^{2}/c^{2}}}{1-\vec{n}\cdot\vec{v}_{\rm em}/c}, \label{eq:zetaD}\\
\zeta_G &=&\frac{1}{\sqrt{|g_{tt}|}}, \label{eq:zetaG}
\end{eqnarray}
with $g_{tt}$ evaluated from Eq.~\eqref{eq:metric} along the worldline of the star and $\vec{n}\cdot\vec{v}_{\rm em}$ the velocity component along the line of sight. Combining the two gives the total relativistic frequency shift
\begin{equation} \label{eq:zetatot}
\zeta=\zeta_D\cdot\zeta_G-1.
\end{equation}
Furthermore, since the velocity frame itself may move along the line of sight, the full observed
radial velocity should also be corrected as \cite{Do:2019txf,GRAVITY:2018ofz}
\begin{equation} \label{eq:VR}
V_R=c\,\zeta+v_{z0} .
\end{equation}
It is noted that additional relativistic effects include the Shapiro delay, gravitational light bending, and frame dragging caused by the spin of Sgr~A*, such as Lense--Thirring precession of the stellar orbit  also modifies photon propagation. The impact of these effects on S2 astrometry and spectroscopy depends on the observational precision and monitoring baseline. But at the current level of observation, it is sufficient to retain the R\o mer delay and pericenter advance and neglects these additional propagation and spin effects \cite{Grould:2017bsw}.

In Fig.~\ref{fig:OBC}, we show and compare the resulting orbits and radial velocities with the data, for the Schwarzschild black hole and for the three dark matter halo models. It is explicit that the four curves are difficult to tell apart on the scale of the plot, meaning that the halo modifies the trajectory only weakly. Thus, a simultaneous fit of the positions, the velocities and the precession is the appropriate way to constrain the model parameters.

\section{Simulation and results}\label{sec:results}

By now, the fifteen parameters can now be estimated from the data. We use the \textsf{emcee} package  \cite{Foreman-Mackey:2012any}, which is well suited to a fifteen-dimensional posterior with strong internal correlations. The walkers are initialised close to the published orbit of S2. The full set of parameters involved in the analysis is
\begin{equation} \label{eq:parameter}
\Theta=\Big\{ M,R_{0},a,e,\iota,\omega,\Omega,t_{\rm apo}, x_0, y_{0}, v_{x0}, v_{y0}, v_{z0}, r_s/M, \rho_s M^2 \Big\}.
\end{equation}
The first two, $M$ and $R_0$, are the mass of Sgr A* and its distance from the Earth. The next six, $a$, $e$, $\iota$, $\omega$, $\Omega$ and $t_{\rm apo}$, are the Keplerian elements of the orbit, the semi-major axis being given in angular units. The five parameters $x_0$, $y_0$, $v_{x0}$, $v_{y0}$ and $v_{z0}$ describe the astrometric zero-point offsets and linear drifts of the reference frame. The last two, $r_s/M$ and $\rho_sM^2$, are the dimensionless parameters of related  halo, of which the constraint are the main target of the current work.

\subsection{Bayesian analysis}
\begin{table}[h]
\centering
\setlength{\tabcolsep}{10pt}
\renewcommand{\arraystretch}{1.2}
\begin{tabular}{|l|c|c|}
\hline
Parameter & \multicolumn{2}{c|}{Gaussian prior} \\
\cline{2-3}
 & $\mu$ & $\sigma$ \\
\hline
$a$ (mas)        & 125.058   & 0.041     \\
$e$              & 0.884649  & 0.000066  \\
$M$ ($10^6 M_{\odot}$) & 4.261    & 0.0012   \\
$t_{\text{apo}}$ (yr)       & 1994.2913 & 0.0016   \\
$x_0$ (mas)      & -0.9     & 0.14     \\
$y_0$ (mas)      & 0.07     & 0.12     \\
$\iota$ ($^\circ$)   & 134.567  & 0.033    \\
$\omega$ ($^\circ$)  & 66.263   & 0.031    \\
$\Omega$ ($^\circ$)  & 228.171  & 0.031    \\
$R_0$ (pc)          & 8246.7   & 9.3   \\
$v_{x0}$ (mas/yr)   & 0.080    & 0.010    \\
$v_{y0}$ (mas/yr)   & 0.0341   & 0.00096  \\
$v_{z0}$ (km/s)     & -1.6     & 1.4      \\
\hline
$r_s/M$ & \multicolumn{2}{c|}{Uniform prior within $[0,0.9]$} \\
$\rho_s M^2$ & \multicolumn{2}{c|}{Uniform prior within $[0,0.9]$} \\
\hline
\end{tabular}
\caption{Priors used for our MCMC analysis. The Gaussian prior data are mainly taken from \cite{GRAVITY:2020gka}. The parameters $r_s/M$ and $\rho_s M^2$ use uniform priors within $[0,0.9]$. Here $pc$ is the parsec, $mas$ is the milliarcsecond, $^\circ$ is the degree, and $yr$ is the year.}
\label{tab:table01}
\end{table}

The aforementioned three sets of data will be fitted together, and the total log-likelihood is their sum,
\begin{equation} \label{eq:likelihood}
\log\mathcal{L}=\log\mathcal{L}_{\rm AP} +\log\mathcal{L}_{\rm VR} +\log\mathcal{L}_{\Delta\varphi},
\end{equation}
with
\begin{eqnarray}
\log\mathcal{L}_{\rm AP} &=& -\frac{1}{2}\sum_{i}\left[  \frac{\bigl(X_i^{\rm obs}-X_i^{\rm the}\bigr)^2} {\bigl(\sigma_{X,i}^{\rm obs}\bigr)^{2}} +\frac{\bigl(Y_i^{\rm obs}-Y_i^{\rm the}\bigr)^2} {\bigl(\sigma_{Y,i}^{\rm obs}\bigr)^{2}}\right], \label{eq:LAP}\\
\log\mathcal{L}_{\rm VR} &=& -\frac{1}{2}\sum_{i} \frac{\bigl(V_{R,i}^{\rm obs}-V_{R,i}^{\rm the}\bigr)^2} {\bigl(\sigma_{V,i}^{\rm obs}\bigr)^{2}}, \label{eq:LVR}\\
\log\mathcal{L}_{\Delta\varphi} &=& -\frac{1}{2}\, \frac{\bigl(\fsp^{\rm obs}-\fsp^{\rm the}\bigr)^2} {\bigl(\sigma_{\fsp}^{\rm obs}\bigr)^{2}}. \label{eq:Lfsp}
\end{eqnarray}
Here the quantities with \textit{obs} denote the observed data of the S2 star at the $i-$th epoch while those labeled by \textit{the} are the corresponding  theoretical values. Recalling that the theoretically predicted values come from the orbit model of section~\ref{sec:dynamics}, i.e., the positions from the geodesic integration in the dark matter halo background followed by the projection onto the sky, and the radial velocity  together with the term Eq.~\eqref{eq:VR}. The third term compares the measured precession Eq.~\eqref{eq:fsp} with Eqs.~\eqref{eq:fsp0}--\eqref{eq:fsp25}.

It is noted that in Eq.~\eqref{eq:likelihood}, we ignore their cross-covariance and use diagonal astrometric errors. These choices define the statistical approximation and source confusion can also introduce temporal correlations \cite{plewa2018unrecognized}. Related S2 analyzes use compressed precession inputs
with different error prescriptions
\cite{DellaMonica:2023dcw,deMartino:2021daj,Fernandez:2023kro}.
Subsequently, the posterior is
\begin{equation}
 p(\bm\Theta\mid D)\propto\mathcal L(D\mid\bm\Theta)\pi(\bm\Theta).
\end{equation}
The priors in our simulation are listed in Table~\ref{tab:table01}, which are referred to \cite{GRAVITY:2020gka} and  Gaussian priors, with the means and standard deviations given in the $\mu$ and $\sigma$ columns. For the remaining two parameters, $r_s/M$ and $\rho_s M^2$, we use uniform priors over $[0,0.9]$, which are physically reasonable according to Fig.~\ref{fig:constrainf_sp}.

\subsection{Results}

We list the best-fit values and $1\sigma$ uncertainties of the fifteen model parameters for the three Dehnen-type dark matter halos in Table~\ref{tab:table02}, and the full corner plots are present in Appendix \ref{app:corners}.  

The $1\sigma$ upper limits of the dark matter halo parameters, $r_s/M$ and $\rho_s M^2$, are 
\begin{eqnarray}
r_s/M&<&0.236,\quad 0.168,\quad 0.030, \label{eq:rbound}\\
\rho_s M^2&<&0.536,\quad 0.578,\quad 0.004, \label{eq:rhobound}
\end{eqnarray}
in the order of $\gamma=0,1,5/2$, respectively. 
We see that the constraints for $\gamma=0$ and $\gamma=1$ are of the same order of magnitude, whereas the constraint  for $\gamma=5/2$ is much tighter. This result can be directly understood from two aspects. On one hand, in Eqs.~\eqref{eq:fsp0} and~\eqref{eq:fsp1}, the dark matter corrections to $\fsp$ for the $(1,4,0)$ and $(1,4,1)$ models are $\frac{2\pi}{3}\rho_0 r_0^3(2-r_0)$ and $\frac{2\pi}{3}\rho_0 r_0^3(3-r_0)$, respectively, which scale as $\rho_0 r_0^3$. However, for the $(1,4,5/2)$ model in Eq.~\eqref{eq:fsp25}, the dark matter correction scales as $\rho_0 r_0^2$. Therefore,  the latter case produces a larger deviation in $\fsp$ at small $r_0$ than the other two models. On the other hand, from Fig.~\ref{fig:constrainf_sp}, it is obvious that  the roughly predicted region in $(r_s/M, \rho_s M^2)$ space for $\gamma=5/2$ is  much narrower than the other two cases.

The central mass of Sgr~A$^{*}$ is estimated as $M/(10^{6}M_{\odot})= 4.23^{+0.08}_{-0.07} , 4.24 \pm 0.07$ and $4.27 \pm 0.07$,  while the corresponding  distance to the Galactic Centre are $R_0 = 8133.66^{+102.51}_{-91.07}\,$pc, $8123.60^{+97.72}_{-89.82}\,$pc, and $8119.24^{+80.71}_{-81.43}\,$pc for the models with $\gamma=0,~1,~5/2$, respectively. It is noted that for all three halo profiles, the variations in both $M$ and $R_0$ are within their respective $1\sigma$ uncertainties ($\sim 0.07 \times 10^{6}M_{\odot}$ and $\sim 80$--$100\,$pc), and the fitted values remain in good agreement with the GRAVITY Collaboration reference values listed in Table~\ref{tab:table01} \cite{GRAVITY:2020gka}. This indicates that introducing the Dehnen-type dark matter halo  does not noticeably shift the fundamental mass and distance estimates of Sgr~A$^{*}$, as expected.

\begin{table}[h]
\centering
\setlength{\tabcolsep}{6pt}
\renewcommand{\arraystretch}{1.2}
\begin{tabular}{|l|c|c|c|}
\hline
\multirow{2}{*}{Parameter} & \multicolumn{3}{c|}{Best-fit value} \\
\cline{2-4}
 & $\gamma=0$ & $\gamma=1$ & $\gamma=5/2$ \\
\hline
$a$ (mas)                  & $127.13^{+0.86}_{-0.87}$  & $127.21^{+0.93}_{-0.88}$ & $127.34^{+0.85}_{-0.81}$ \\
$e$                        & $0.89 \pm 0$             & $0.89 \pm 0$ & $0.89 \pm 0$ \\
$M$ ($10^6 M_{\odot}$)     & $4.23 ^{+0.08}_{-0.07}$          & $4.24 \pm 0.07$ & $4.27 \pm 0.07$ \\
$t_{{\rm apo}}$ (yr)      & $1994.29 \pm 0.01$       & $1994.29 \pm 0.01$ & $1994.29 \pm 0.01$ \\
$\mathsf{x}_0$ (mas)       & $0.93^{+0.19}_{-0.20}$          & $0.93 ^{+0.21}_{-0.20}$ & $0.99 \pm 0.20$ \\
$\mathsf{y}_0$ (mas)       & $-2.48 ^{+0.49}_{-0.53}$         & $-2.53 \pm 0.50$ & $-2.77 \pm 0.48$ \\
$\iota$ ($^\circ$)        & $134.01^{+0.30}_{-0.32}$        & $133.99^{+0.31}_{-0.33}$ & $133.92 \pm 0.30$ \\
$\omega$ ($^\circ$)        & $65.28 \pm 0.38$         & $65.29 \pm 0.37$ & $65.06 \pm 0.37$ \\
$\Omega$ ($^\circ$)       & $226.76^{+0.38}_{-0.40}$        & $226.76^{+0.39}_{-0.40}$ & $226.44 \pm 0.41$ \\
$R_0$ (pc)                 & $8133.66^{+102.51}_{-91.07}$ & $8123.60^{+97.72}_{-89.82}$ & $8119.24^{+80.71}_{-81.43}$ \\
$\mathsf{v}_{x0}$ (mas/yr) & $0.08 \pm0.30$          & $0.07 ^{+0.29}_{-0.30}$ & $0.07^{+0.30}_{-0.31}$ \\
$\mathsf{v}_{y0}$ (mas/yr) & $0.10^{+0.22}_{-0.34}$          & $0.05 ^{+0.29}_{-0.31}$ & $0.04^{+0.29}_{-0.30}$ \\
$\mathsf{v}_{z0}$ (km/s)   & $-0.80\pm1.33$         & $-0.80^{+1.38}_{-1.33}$ & $-1.55 \pm 1.38$ \\
$r_s/M$                    & $<0.236$        & $<0.168$ & $<0.030$ \\
$\rho_s M^2$               & $<0.536$        & $<0.578$ & $<0.004$ \\
\hline
\end{tabular}
\caption{The best-fit values of the parameters of the orbital model of S2 under Gaussian priors for the Dehnen-$(1, 4, \gamma)$ dark matter halo models, with $\gamma = 0, 1,$ and $5/2$, respectively. The uncertainties are given at the $1\sigma$ level.}
\label{tab:table02}
\end{table}

Moreover, the exterior $1/r$ term in the first two metrics gives the dynamical mass
$\MK=M+\Mh$. In terms of this mass, Eq.~\eqref{eq:unified} becomes
\begin{equation}
 \Deltaom=\frac{6\pi\MK}{p}-\frac{\pi Q}{\MK p}+\cdots.
 \label{eq:massdegeneracy}
\end{equation}
The positive fixed-$M$ contribution from $\Mh$ is a monopole mass shift.
The coefficient $Q$ controls the first radial-profile correction.
This separation explains the correlation between a compact halo
contribution and the fitted central metric mass.
But, for $\gamma=5/2$, circular geodesics measured in normalized time satisfy
$\Omega_T^2=f'(r)/(2\kappa r)$, giving $\MK=\mu/\kappa$.
The advance is then
\begin{equation}
 \Deltaom=2\pi(\kappa^{-1/2}-1)
 +\frac{6\pi\MK}{\sqrt\kappa p}
 -\frac{\pi Q}{\MK\kappa^{3/2}p}+\cdots.\label{eq:normalized}
\end{equation}
The first term measures the angular deficit of the spatial geometry.

The nominal reference mass converts the limits in Eq.~\eqref{eq:rbound}
to
\begin{equation}
 r_s<0.00992,\quad0.00706,\quad0.00126\ \mathrm{AU}.
 \label{eq:physicalradii}
\end{equation}
The full prior range $r_s<0.9M$ lies below the vacuum horizon scale $2M$,
while the S2 pericenter is about $2.83\times10^3M$.
S2 therefore probes the exterior coefficients of these metrics on scales
far outside $r_s$. The fitted scale radius is unresolved by the stellar
orbit.

We note that the constraints in Eqs.~\eqref{eq:rbound} and \eqref{eq:rhobound} are separate
marginals. A mass or coefficient such as $\Mh$ or $Q$ depends on their joint distribution, which requires paired posterior samples. In Appendix~\ref{app:source}, we gives the 
corresponding retrograde mass precession for the physical Dehnen profile via the weak-field calculation.

\section{Conclusion and discussions}
We have used the orbit of S2 to constrain three Dehnen-inspired black hole metrics associated with inner slopes $\gamma=0$, $1$, and $5/2$. The analytical calculation retains the large eccentricity of S2 and identifies the leading halo contributions to its pericenter advance. A fifteen-parameter Bayesian analysis combines astrometric and spectroscopic observations with the measured Schwarzschild-precession parameter. The marginalized $1\sigma$ upper constraints are
\begin{eqnarray}
r_s/M < 0.236,\quad 0.168,\quad 0.030;\
~~~~~~\rho_s M^2 < 0.536,\quad 0.578,\quad 0.004,
\end{eqnarray}
respectively. The inferred central mass and distance vary by less than their marginal uncertainties across the three metrics. The fitted orbital orientations and reference-frame parameters also have overlapping uncertainties.

The smaller limits for the $\gamma=5/2$ branch reflect its stronger precession response in the adopted parameterization. For $\gamma=0$ and $1$, the leading corrections to $\fsp$ scale as $\rho_0r_0^3$ at small $r_0$. The third metric approaches $\kappa=1-\mathcal A$ at large radius, where $\mathcal A=32\pi\rho_0r_0^2$. Its spatial angular deficit contributes $\Deltaom_{\rm deficit}\simeq\pi\mathcal A$ at linear order, giving $\delta \fsp\simeq\mathcal A p/(6M)$. The large ratio $p/M$ makes S2 sensitive to this contribution. The difference between the three sets of limits therefore follows from their distinct radial and asymptotic geometries.

The exterior expansion also clarifies the role of mass degeneracy. For the two asymptotically flat branches with $\gamma=0, 1$, the leading halo monopole enters through the dynamical mass $\MK =M+\mathcal M_{\rm D}$. At fixed $\MK$, the first profile-dependent contribution to the pericenter advance is $-\pi Q/(\MK p)$, where $Q=4\pi\rho_s r_s^4$. This retrograde term agrees with the compact limit of the mass precession produced by a dilute Dehnen distribution. Throughout the adopted scale-radius prior, S2 lies far outside $r_s$ and primarily probes the exterior metric coefficients. The inner density profile remains unresolved by the stellar orbit. The resulting bounds apply to the specified metric families, whose Einstein-source densities differ from the Dehnen profiles that generate their Newtonian radial forces.

The orbital constraints in this work probe the current geometries on scales much larger than those sampled by black-hole shadows. Previous shadow studies of the same metric families obtained constraints by fixing one halo parameter and varying the other \cite{Jha:2024ltc,Jha:2025xjf,Liang:2025vux}. The present bounds instead follow from marginalization over the joint orbital and halo posterior. The published constraints also use different uncertainty conventions and, in some cases, observations of a different black hole. A quantitative comparison of constraining power therefore requires matched parameter conditioning, probability content, and mass calibration. For the $\gamma=5/2$ branch, the nonflat asymptotic geometry also enters the conversion from photon impact parameters to observed shadow angles \cite{Perlick:2021aok}.

It is noted the separate upper limits on $r_s$ and $\rho_s$ can not determine an upper limit on the halo mass or on the coefficients of the exterior metric, which depend on the correlated combinations of the fitted parameters and should be obtained directly from paired posterior samples. Reporting their joint distributions would provide a closer connection between the statistical bounds and the gravitational field measured by S2 star. Additionally, a joint fit of several S-stars with different orbital sizes and eccentricities could help distinguish the central monopole from profile-dependent corrections by sampling a broader radial range. Combining such an analysis with the Sgr A* shadow, using shared mass and distance parameters and a consistent photon-propagation model, would test the same geometry on stellar-orbit and near-horizon scales.

\begin{acknowledgments}
This work is partly supported by Natural Science Foundation of China under Grants Nos.12375054 and W2433018, and the Postgraduate Research $\&$ Practice Innovation Program of Jiangsu Province under Grant No. $\text{KYCX25\_3924}$. 
\end{acknowledgments}

\appendix
\section{Source densities and the Dehnen mass-precession benchmark}
\label{app:source}

Equation~\eqref{eq:stress} gives
\begin{align}
 \rho_{\mathrm E}^{(0)}&=
 \frac{\rho_sr_s^4(r_s+3r)}{6r^2(r+r_s)^3},\label{eq:rhoeff0}\\
 \rho_{\mathrm E}^{(1)}&=
 \frac{\rho_sr_s^4}{2r^2(r+r_s)^2},\label{eq:rhoeff1}\\
 \rho_{\mathrm E}^{(5/2)}&=
 \frac{2\rho_sr_s^2(r_s+2r)}{r^{5/2}\sqrt{r+r_s}}.
 \label{eq:rhoeff25}
\end{align}
The first two source densities have an algebraic inner $r^{-2}$
dependence and an outer $r^{-4}$ tail. The inner limit characterizes the
functions and the static-source interpretation applies outside the horizon.
For the third branch,  we have 
\begin{equation}
 m_{\mathrm E}^{(5/2)}-M
 =16\pi\rho_sr_s^2\sqrt{r(r+r_s)},\label{eq:mass25}
\end{equation}
which grows linearly at large radius and gives an $r^{-2}$ source tail.

A general spherical matter model has two metric functions,
\begin{equation}
 \dd s^2=-A(r)\dd t^2+\frac{\dd r^2}{1-2m(r)/r}+r^2\dd\Omega^2.
\end{equation}
Its field equations include
\begin{equation}
 m'=4\pi r^2\rho,\qquad
 \frac{A'}A=\frac{2[m+4\pi r^3p_r]}{r(r-2m)}.\label{eq:einstein}
\end{equation}
The density and pressure together determine the redshift potential
\cite{Cardoso:2021wlq}.

For a genuine Dehnen mass distribution that perturbs a dominant Kepler mass $\MK$, the first-order Gauss equation yields \cite{Adkins:2007et}
\begin{equation}
 \Deltaom_{\mathrm D}
 =\frac1{\MK e}\int_0^{2\pi}
 M_{\mathrm D}[<r(\chi)]\cos\chi\,\dd\chi.\label{eq:gauss}
\end{equation}
A constant interior mass integrates to zero.
By defining $n=3-\gamma$, $B=p+r_s$, and $z=(er_s/B)^2$.
we can reduce the above integral as
\begin{equation}
 \Deltaom_{\mathrm D}=-\pi n\frac{\Mh}{\MK}
 \frac{r_sp^n}{B^{n+1}}
 {}_2F_1\left(\frac{n+1}{2},\frac{n+2}{2};2;z\right).
 \label{eq:physicalprecession}
\end{equation}
For positive density, $0\leq\gamma<3$, and $0\leq e<1$, the mass precession is retrograde. The circular case is the continuous $e\to0$ limit. Eq.~\eqref{eq:physicalprecession} is valid at arbitrary $r_s/p$ for a dilute perturbation and its compact limit is
\begin{equation}
 \Deltaom_{\mathrm D}=-\frac{4\pi^2\rho_sr_s^4}{\MK p}+\cdots,
\end{equation}
which is in agreement with the radial-profile term in
Eq.~\eqref{eq:unified}. 

\section{Posterior distributions}\label{app:corners}

\begin{figure} [h]
{\centering
\includegraphics[width=5in]{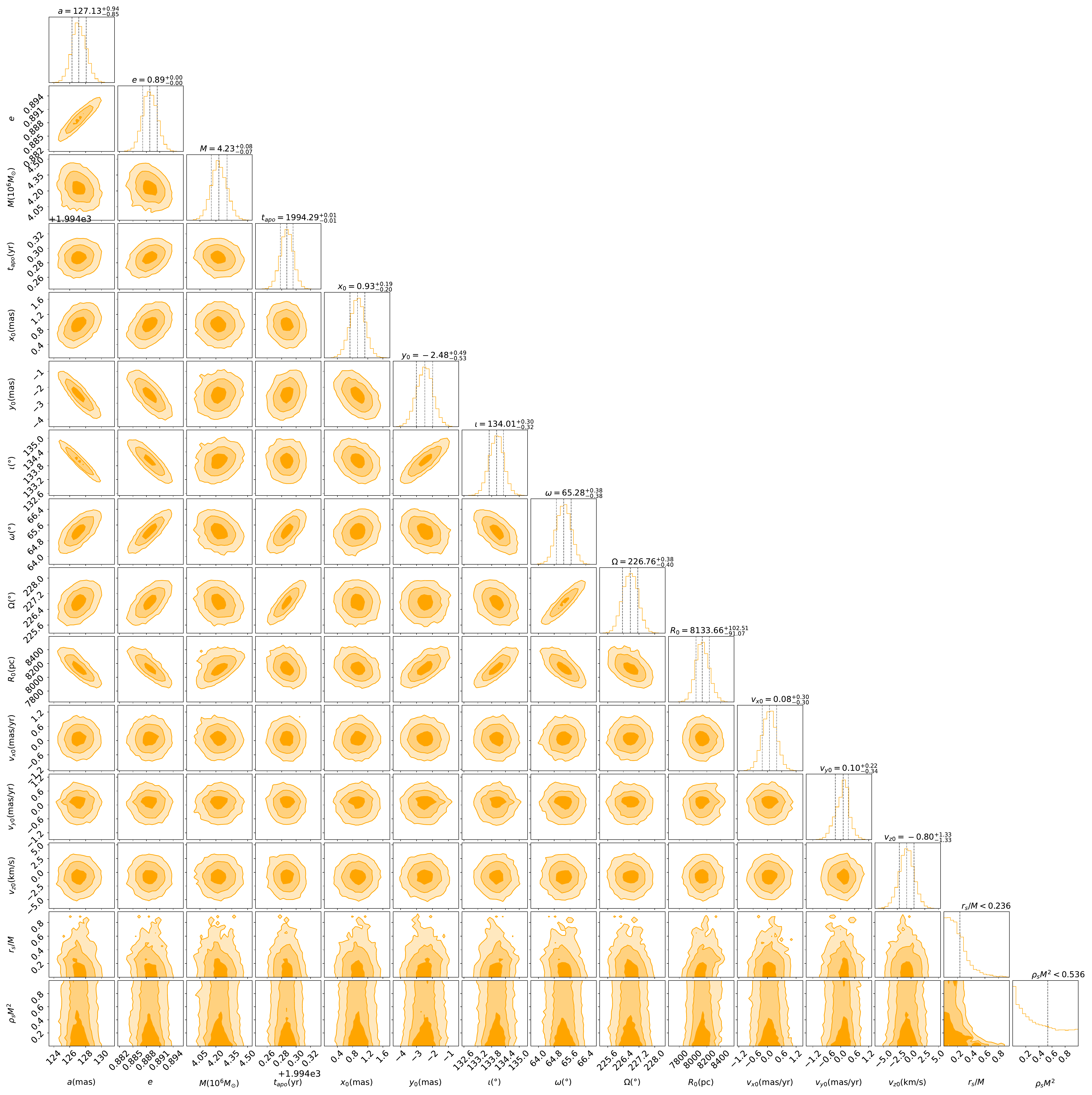} \hspace{0.5cm}
   \caption{ The figure shows the posterior distributions of the orbital parameters for S2 orbiting a black hole surrounded by a Dehnen-(1, 4, 0) type DM halo. The dashed lines mark the best-fit values and 1$\sigma$ confidence levels.}   \label{fig:140_Gau}}
\end{figure}

\begin{figure} [h]
{\centering
\includegraphics[width=5in]{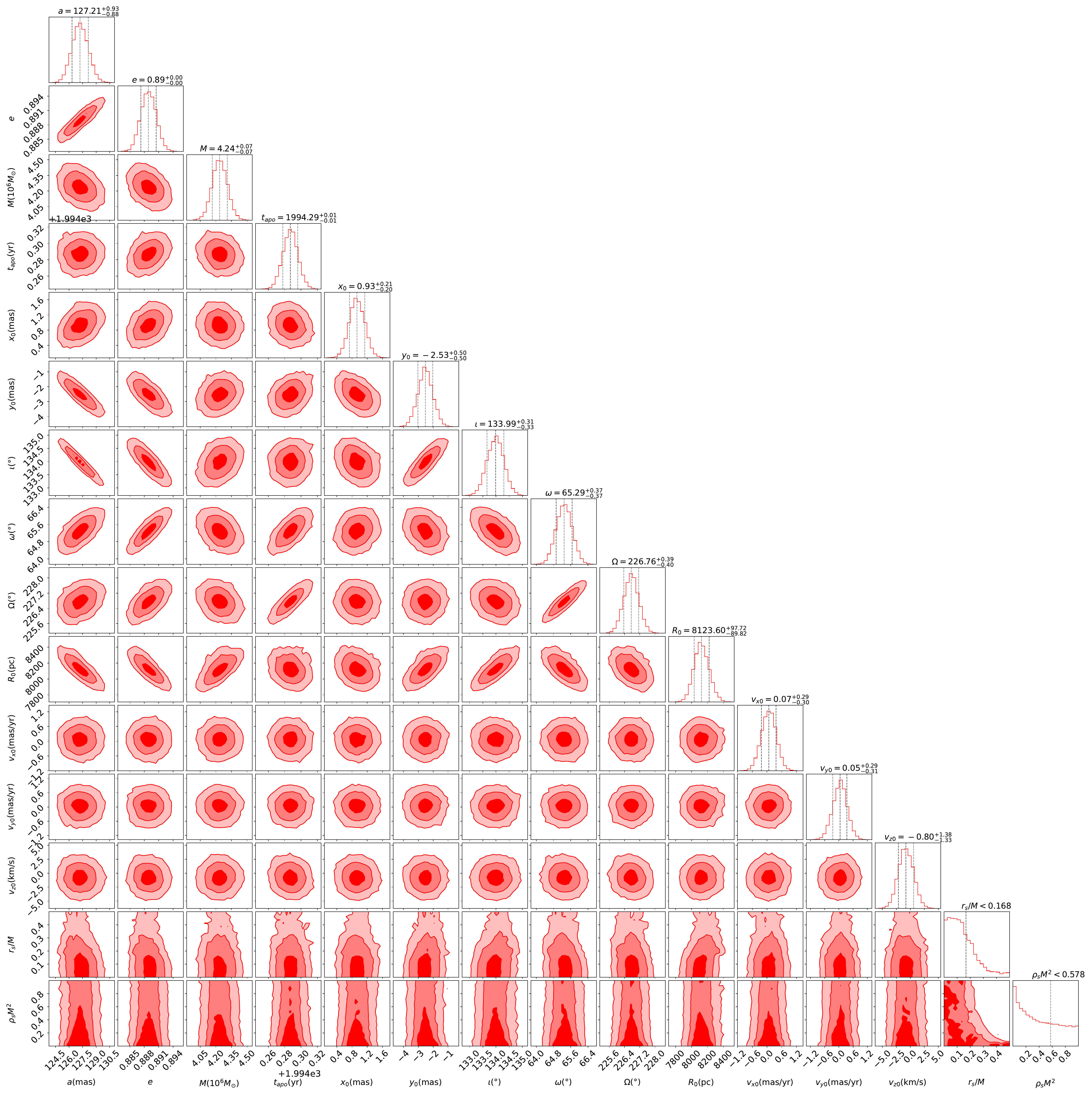} \hspace{0.5cm}
   \caption{ Posterior distributions for the $\gamma=1$ metric.
 The panel layout follows Fig.~\ref{fig:140_Gau}.}   \label{fig:141_Gau}}
\end{figure}

\begin{figure} [h]
{\centering
\includegraphics[width=5in]{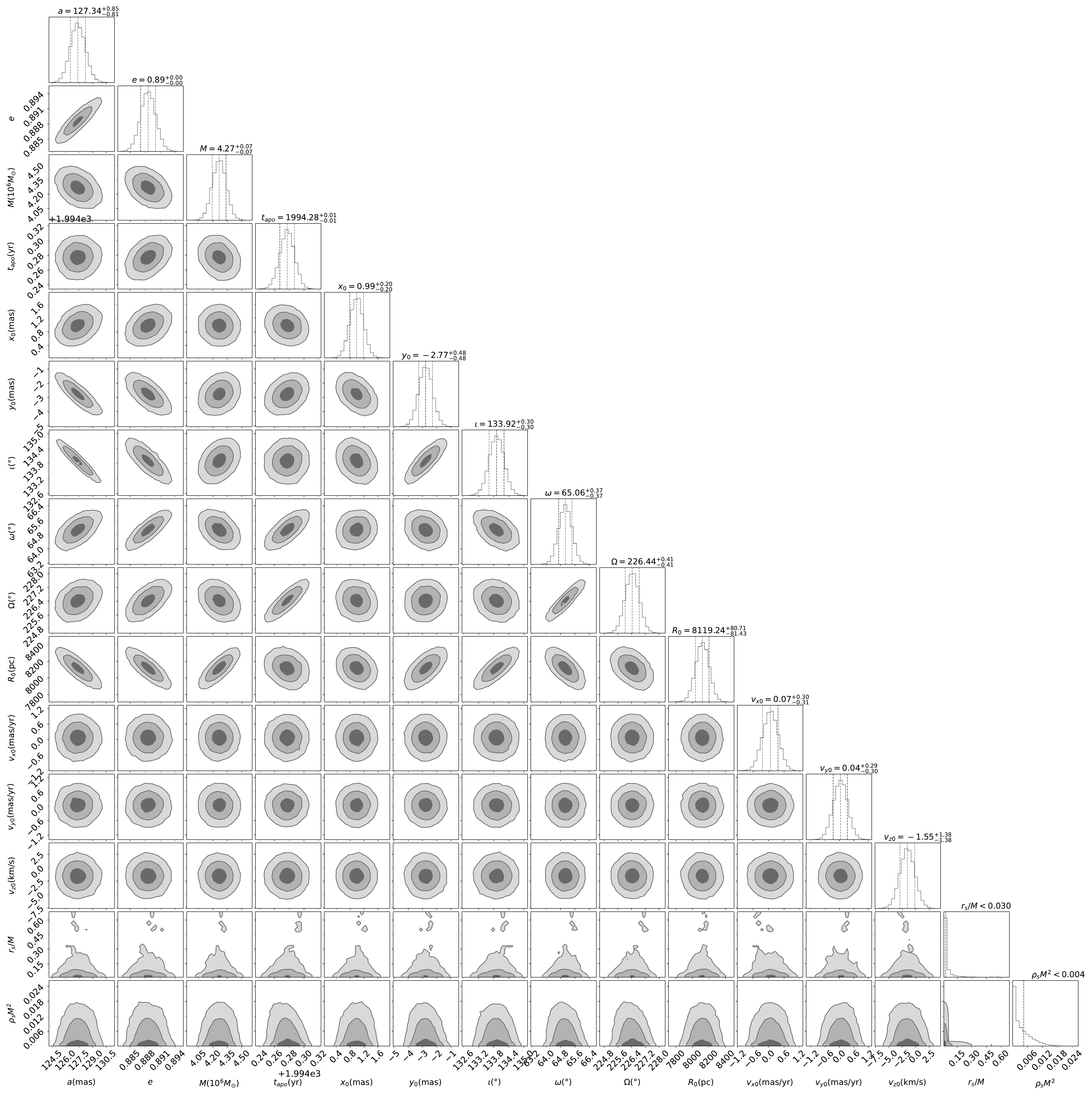} \hspace{0.5cm}
   \caption{ Posterior distributions for the $\gamma=5/2$ metric.
 The panel layout follows Fig.~\ref{fig:140_Gau}.}     \label{fig:1452_Gau}}
\end{figure}

Figs.~\ref{fig:140_Gau}--\ref{fig:1452_Gau} show the joint and marginal
distributions of the fifteen fitted parameters for the three metric
families. The parameter order is the same in all figures.

\bibliography{ref}
\bibliographystyle{utphys}

\end{document}